\documentclass[Ap&SS, manuscript]{copernicus}

\usepackage{setspace}
\usepackage{lineno}
\usepackage{graphicx}
\usepackage{color}
\usepackage{caption}
\usepackage{subcaption}

\begin{document}

\setpagewiselinenumbers
\modulolinenumbers[100]
\linenumbers

\title{Statistical Analysis of I Stokes Parameter of Millisecond Pulsars}


\Author{Hossein}{Panahi}
\Author{Issa}{Eghdami}
\Author{Reza}{Monadi}
\affil{Department of Physics, University of Guilan, Rasht 41635-1914, Iran}


\runningtitle{Gaussian Anisotropy In Strange Quark StarsTEXT}
\runningauthor{Hossein Panahi, Issa Eghdami, and Reza Monadi}

\correspondence{H. Panahi (t-panahi@guilan.ac.ir)}

\received{199}
\pubdiscuss{} 
\revised{} \accepted{} \published{}


\firstpage{1}

\maketitle

\begin{abstract}
Using Detrended Fluctuation Analysis (DFA) and box counting method, we test spacial correlation and fractality of Polarization Pulse Profiles (PPPs) of 24 millisecond pulsars (MSPs) which were observed in Parkes Pulsar Timing Array (PPTA) project. DFA analysis indicates that MSPs' PPPs are persistent and the results of box counting method confirm the fractality in the majority of PPPs.  A Kolmogorov-Smirnov test indicates that isolated MSPs have more complex PPPs than binary ones. Then we apply our analysis on a random sample of normal pulsars. Comparing the results of our analysis on MSPs and normal pulsars shows that MSPs have more complex PPPs which is resulted from smaller angular half-width of the emission cone and more peaks in MSPs PPPs. On the other hand, high values of Hurst exponent in MSPs confirm compact emission regions in these pulsars.
\end{abstract}

\section{Introduction}
Millisecond pulsars (MSPs) are a subgroup of pulsars which are rapidly spinning, so that their rotational period is within the range of milliseconds. In theory MSPs are old and weakly magnetized neutron stars which are recycled because of accretion of matter from a companion star (\cite{evolution}). \\
There are 24 MSPs observed by PPTA project and their PPPs released by \cite{PPTA} in three bandwidths (10\,cm, 20\,cm, 50\,cm) which this quality encouraged us to use this sample of MSPs.
Studying Stokes polarization parameters defined by George Gabriel Stokes (\cite{Chandrasekhar}) have yielded valuable information about magnetic field and pulse emission mechanism of pulsars. In astronomical conventions Stokes parameters are denoted by I (total intensity), $L=\sqrt{Q^{2}+U^{2}}$ (total linear polarization) and V (circular polarization). The PSR/IEEE convention is the standard implemented by PSRCHIVE software (\cite{stokes}) which is utilized to fold and export the data of PSRFITS files in this work.
The mean pulse profile of pulsar is a stable property of pulsars studied by \cite{Rankin}, \cite{lyne}, \cite{manchester2} , and others.
There are two major ideas to explain the shape of pulsars' PPPs. In the first one, introduced by Rankin and others, emission beams are divided into core and conal components (\cite{Rankin}). Another one, which is known as "patchy model", was represented in (\cite{manchester2}) by introducing window and source functions. Window function, the common features carried in all pulsars, depends on radio frequency and pulsar's period. On the other hand, the source function is a unique feature of each pulsar and represents the random distribution of emission sources.
The core and cone emission idea can be used to explain symmetric Pulse Profiles (PPs), while "patchy" model explains the shape of asymmetric ones (\cite{Karastergiou}). In order to study the shape and complexity of pulse profiles, \cite{Karastergiou} introduced a combination of patchy and core models. This model indicates that the emission regions of younger pulsars are in a narrow range in high altitudes and then the pulse profiles are simpler in number of peaks in comparison with old pulsars. It is often believed that the radio emission mechanism of normal pulsars and MSPs  are the same  (\cite{KramerI}), however there are differences in properties of observed pulse profiles for these two groups. Taking into account this assumption, we compare the statistical properties of two samples (D and $\alpha$ exponent) and also their correlation with pulsars characteristic parameters.\\
Although pulse profile is a fingerprint of pulsars, there is not enough information about the emission mechanism in pulsars. Hence studying the statistical properties of PPPs can reveal valuable information about the nature of pulse emission. Here we used two statistical method, namely DFA and box counting. In 1951 Edwin Hurst invented a rescaled range analysis ($R/S$) during studying long term memory of Nile river which is not a reliable method specially in very short time series (\cite{Racine}). In (\cite{Peng}) DFA analysis proposed as a reliable method for studying stationary and non-stationary time series and we implement it to study PPPs long term memory. In addition box counting method can be used as a quantitative measure for complexity, which characterizes the scaling properties of a pattern and its self-similarity. In fact the complex behavior in intensity of pulse profiles  and the existence of smaller subpulses, micropulses, and even nanopulses alongside main pulses  motivated us to study the fractality of PPPs (\cite{hankin}). We expect, the more peaks and  the more fluctuations a particular PPP has, the higher fractal dimension in that PPP will be appeared. Thus box counting and DFA methods can respectively lead to local and global statistical information about pulsars' PPPs. The main goal of this work is manifesting the statistical properties of emission regions in MSPs' PPPs. We have two main ideas to test:
\begin{itemize}
\item {\textit{PPP's fluctuations are spaced correlatively.}}
\item{\textit{Pulse profiles have fractal characteristics.}}
\end{itemize}
Box counting and DFA method were performed for testing these two ideas. Then to make a comparison with normal pulsars we apply our analysis to a random sample of 24 normal ones. We use the box counting analysis to investigate the fractality and therefore the complexity of pulses quantitatively. On the other hand, DFA method leads us to the Hurst exponent which explains the correlation of a signal fluctuations and therefore the correlation of emission regions.
In \S \ref{s2} we will represent DFA and box counting analysis methods. In \S \ref{pps} we will test our analyses on some simulated pulse profiles. Finally we will illustrate our results and discuss them in \S \ref{res}.

\section{Analysis methods}\label{s2}
In the following subsections we are intended to utilize  statistical methods to study the PPPs: detrended fluctuation and fractal analysis. Then we investigate relationship between the outputs of these methods and pulsars characteristic parameters in various bandwidths of Stokes parameter I.

\subsection{Detrended Fluctuation Analysis}
In this work we apply  DFA analysis  which is a standard method for determining global properties of a system. The advantage of this method is that we do not have to make any assumption about the stationarity or non-stationarity of the signal, hence it can be used for both of stationary and non-stationary processes (\cite{Hardstone}). On the other hand since the shape of pulse profiles depends on latitude of the magnetic axis and the observer's line of sight (\cite{manchester2}), the detrending procedure of this analysis will be an advantage for illuminating the nature of a typical PPPs' source.
To study the long term memory of PPPs we use the Hurst exponent $H$, defined by Edwin Hurst, which is a dimensionless estimator for the self similarity and long term memory. A value of $H>0.5$ means positive long term memory or persistent time series, while $H<0.5$ implies a negative long term memory or anti-persistence and finally $H=0.5$ refers to a completely independent time series.
DFA method introduced by \cite{Peng} can be explained in the following steps for a time series of length N (\cite{Weron,Hu}):
\begin{enumerate}
\item Divide time series $X_{i},$ $i=1,2,...,N$ into $N_{w}$ windows of length $\tau$.
\item Create mean adjusted cumulative subseries to avoid the overestimation of DFA method at small scales (\cite{Chen}):
\begin{equation}
Y(j,\nu) = \sum_{i=1}^{j} ( X(i) - < X > ),
\label{cumulative}
\end{equation}
where $1\leq j \leq \tau$ for each period of $\tau$ and $\nu$ is the number of window.
\item Remove the trend of each subseries by a linear least square fit ($Y_{fit}$):
\begin{equation}
Z(j,\nu) = Y(j,\nu)-Y_{fit}(j,\nu).
\label{detrend}
\end{equation}
$Y_{fit}$ can be a polynomial fit of order "$n$" and the corresponding DFA method is called DFAn (\cite{Rypdal}). In current work we use linear fit or DFA1.
\item Calculate the root mean square fluctuation of detrended cumulative series:
\begin{equation}
F^{2}(\tau,\nu) =  { \frac{1}{\tau}\sum_{j=1}^{\tau} [Z(j,\nu)]^{2} },
\label{STD1}
\end{equation}

\begin{equation}
F(\tau) = \sqrt { \frac{1}{N_{w}}\sum_{\nu=1}^{N_{w}} [F(\tau,\nu)]^{2} }.
\label{STD}
\end{equation}
\item Repeat steps (i) through (iv) for different values of $\tau$. There is a power-law scaling relation between resulted root mean square fluctuation $F(\tau)$ and window length $\tau$ as:
\begin{equation}
F(\tau) \sim \tau^{\alpha}.
\label{power}
\end{equation}
\end{enumerate}

It is easy to see that the detrending procedure in the windows of length 2 ($\tau = 2$) leads to same result for any data set and by construction, DFA1 is defined for $\tau \geq 3$. On the other hand, this method is unreliable for very large windows ($\tau \geq \frac{N}{4}$) because the number of windows ($N_{w}$) becomes very small (\cite{mfDFA}).
The value of $0<\alpha<1$ implies a stationary process and can be modeled by a fractional Gaussian noise (fGn) with $H=\alpha$. In the case of $\alpha>1$ the process is non-stationary and can be modeled by a fractional Brownian motion (fBm) with $H=\alpha - 1$ (\cite{movahed,Hardstone}). As a result although the DFA method is well-known for analyzing time series, the calculated Hurst exponent can be defined in spatial patterns such as rough surfaces (\cite{japoni}).
\subsection{Fractal Analysis}
A fractal is a shape consisted of parts similar to the whole in some way. In other words a given set has fractal properties if its Hausdorff-Besicovitch dimension $D$ strictly exceeds the topological dimension $D_{T}$. The fractal dimension  can be determined using  box counting method which is widely used  to study the fractal dimension of 1, 2, and 3 dimensional patterns. In this procedure, the pattern of a particular set is covered entirely  by $N_{b}$ boxes of $\xi$ size. The number of boxes which cover the whole pattern is a function of box size as follows (\cite{Feder,japoni}):
\begin{equation}
N_{b}(\xi) \sim \xi^{-D},
\label{box}
\end{equation}
where D can be evaluated by a linear fit on a log-log plot of number of boxes $N_{b}(\xi)$ versus box size $\xi$. Clearly, using bigger (smaller) boxes requires fewer (more) number of them to cover the whole pattern. In effect, box counting method is generally used to determine the fractal dimension of images and rough surfaces, but it also can be implemented in signal analysis (\cite{fractal_signal}).

\section{Pulse Profile Study} \label{pps}
\subsection{Simulated pulse profiles} \label{sim}
\begin{figure}
        \centering
          \begin{tabular}{cc}

        \begin{subfigure}[b]{0.5\textwidth}
                \includegraphics[width=\textwidth]{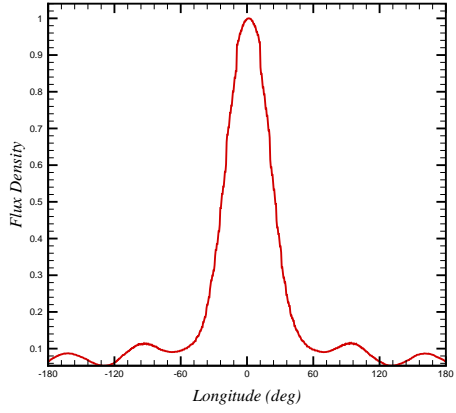}
                \caption{$\rho=17^{\circ}$, $\alpha=1.78\pm0.09$ , $D=1.14\pm0.07$.}
                \label{fig:17}
        \end{subfigure}%

        \begin{subfigure}[b]{0.5\textwidth}
                \includegraphics[width=\textwidth]{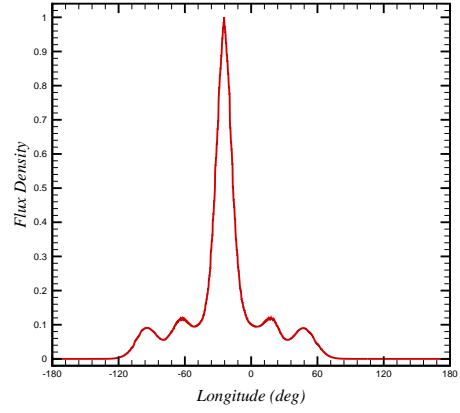}
                \caption{$\rho=34^{\circ}$, $\alpha=1.76\pm0.08$, $D=1.08\pm0.09$.}
                \label{fig:34}
        \end{subfigure}
\\
        \begin{subfigure}[b]{0.5\textwidth}
                \includegraphics[width=\textwidth]{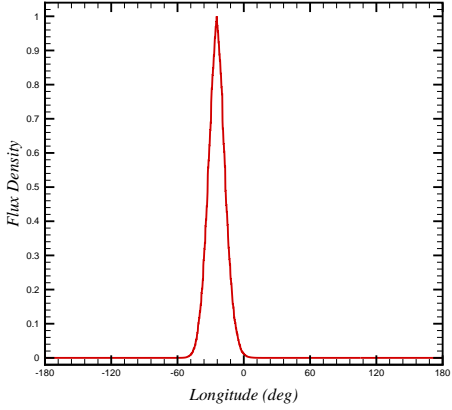}
                \caption{$\rho=34^{\circ}$, $\alpha=1.77\pm0.09$, $D=1.06\pm0.07$ without sub-pulses.}
                \label{fig:34_NoSubpulse}
        \end{subfigure}

        \begin{subfigure}[b]{0.5\textwidth}
                \includegraphics[width=\textwidth]{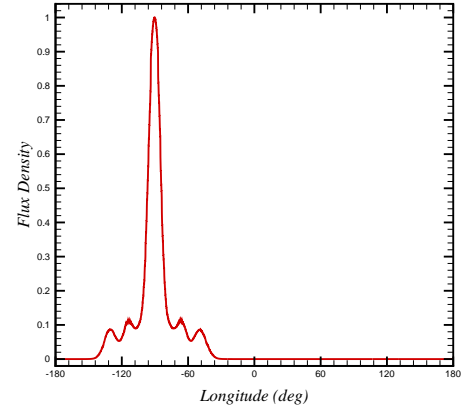}
                \caption{$\rho=68^{\circ}$, $\alpha=1.72\pm0.09$, $D=1.05\pm0.08$.}
                \label{fig:68}
        \end{subfigure}

          \end{tabular}
        \caption{Simulated PPs of a MSP at 1.4 GHz with the same inclination and impact angles $( \theta=20^{\circ}, \beta=-4^{\circ})$. }
        \label{figure:simulation1}
\end{figure}

In this work we have introduced fractal dimension as a new measurement of complexity. A simple simulation of pulse profiles can reveal the meaning of complexity and Hurst exponent in these pulse profiles. We have used core and conal model of a relativistic emission to simulate pulse profile of a MSP (\cite{Gil}). For certain values of inclination angle $(\theta=20^{\circ})$ and impact angle $(\beta=-4^{\circ})$, by varying angular half-width of the emission cone ($\rho$) in a same core and cone emission setup we have concluded that:
\begin{itemize}
\item{Fractal dimension has been decreased by increasing $\rho$. }
\item{$\alpha$ exponent and therefore the Hurst exponent have been decreased by increasing $\rho$. }
\end{itemize}
Our results show the complexity not only depends on the number of peaks and amount of noise, but also on the peaks' height and width  and is eventually related to the size of emission cap (see Fig. \ref{figure:simulation1}). Calculating the Hurst exponent of simulated PPPs reveals that the bigger half-width of the emission cone, the smaller Hurst exponent will be. This trend is reasonable because by increasing polar cap size, more scattered emission regions will participate in PPP. This fact in turn makes PPP more random and finally we get a  Hurst exponent near to 0.5 in the majority of normal pulsars.

\subsection{Fitting on pulse profiles} \label{fit}
\begin{figure}
        \centering
          \begin{tabular}{cc}

        \begin{subfigure}[b]{0.5\textwidth}
                \includegraphics[width=\textwidth]{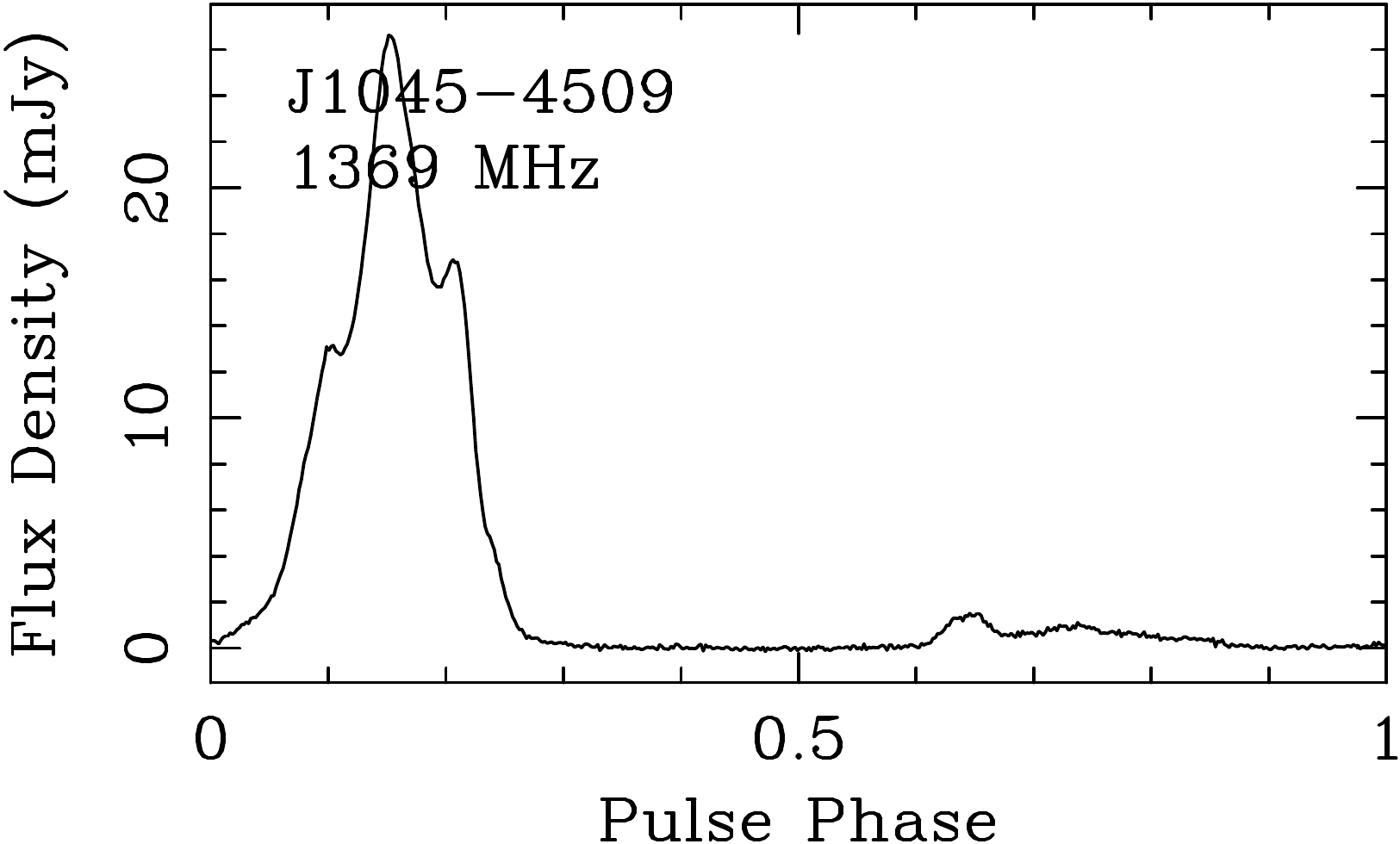}
                \caption{J1045-4509, $D=1.13 \pm 0.07$, Number of observation epoches: 137.}
                \label{fig:1}
        \end{subfigure}%
\\
        \begin{subfigure}[b]{0.5\textwidth}
                \includegraphics[width=\textwidth]{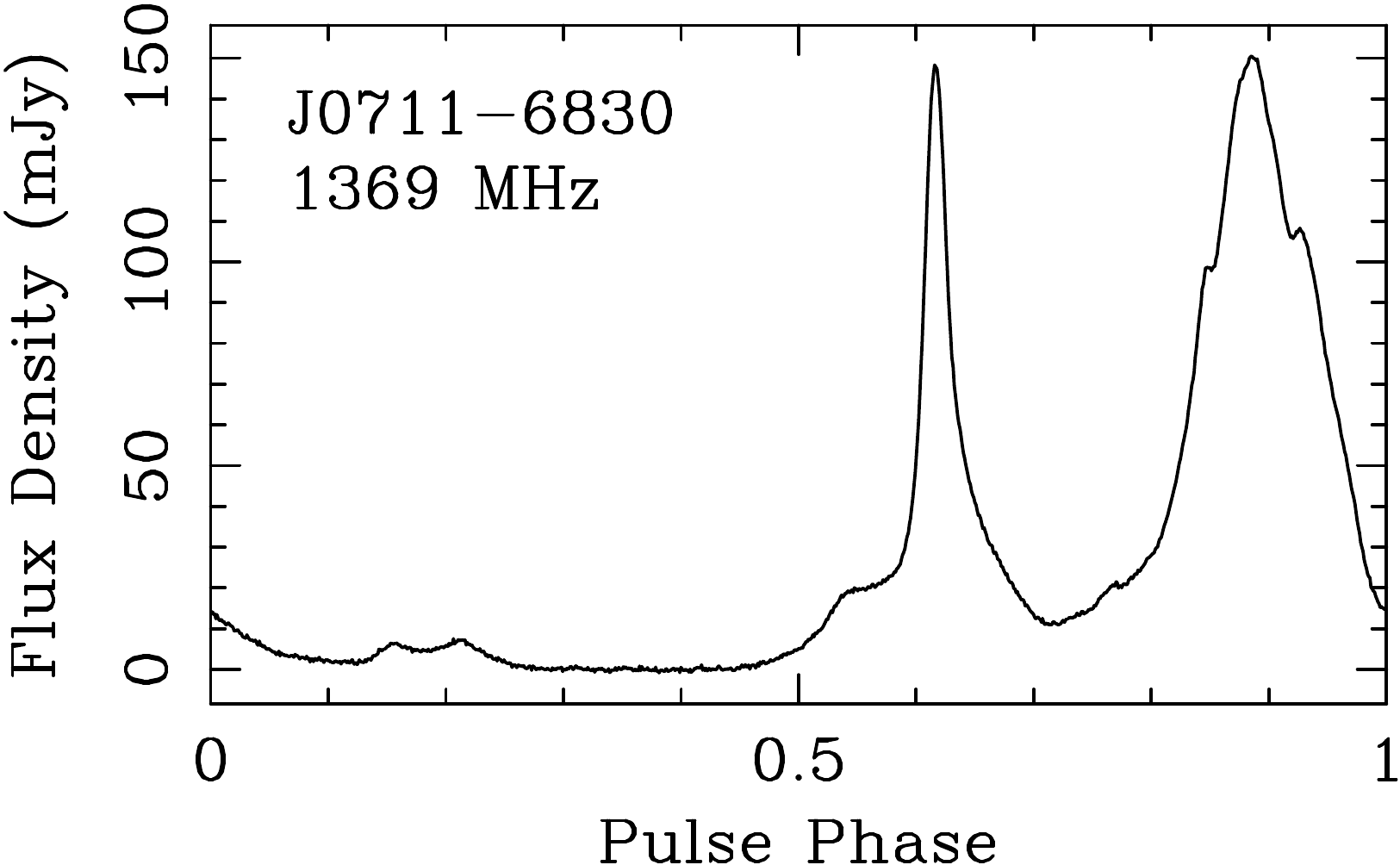}
                \caption{J0711-6830, $D=1.23 \pm 0.08$, Number of observation epoches: 161.}
                \label{fig:2}
        \end{subfigure}
\\
        \begin{subfigure}[b]{0.5\textwidth}
                \includegraphics[width=\textwidth]{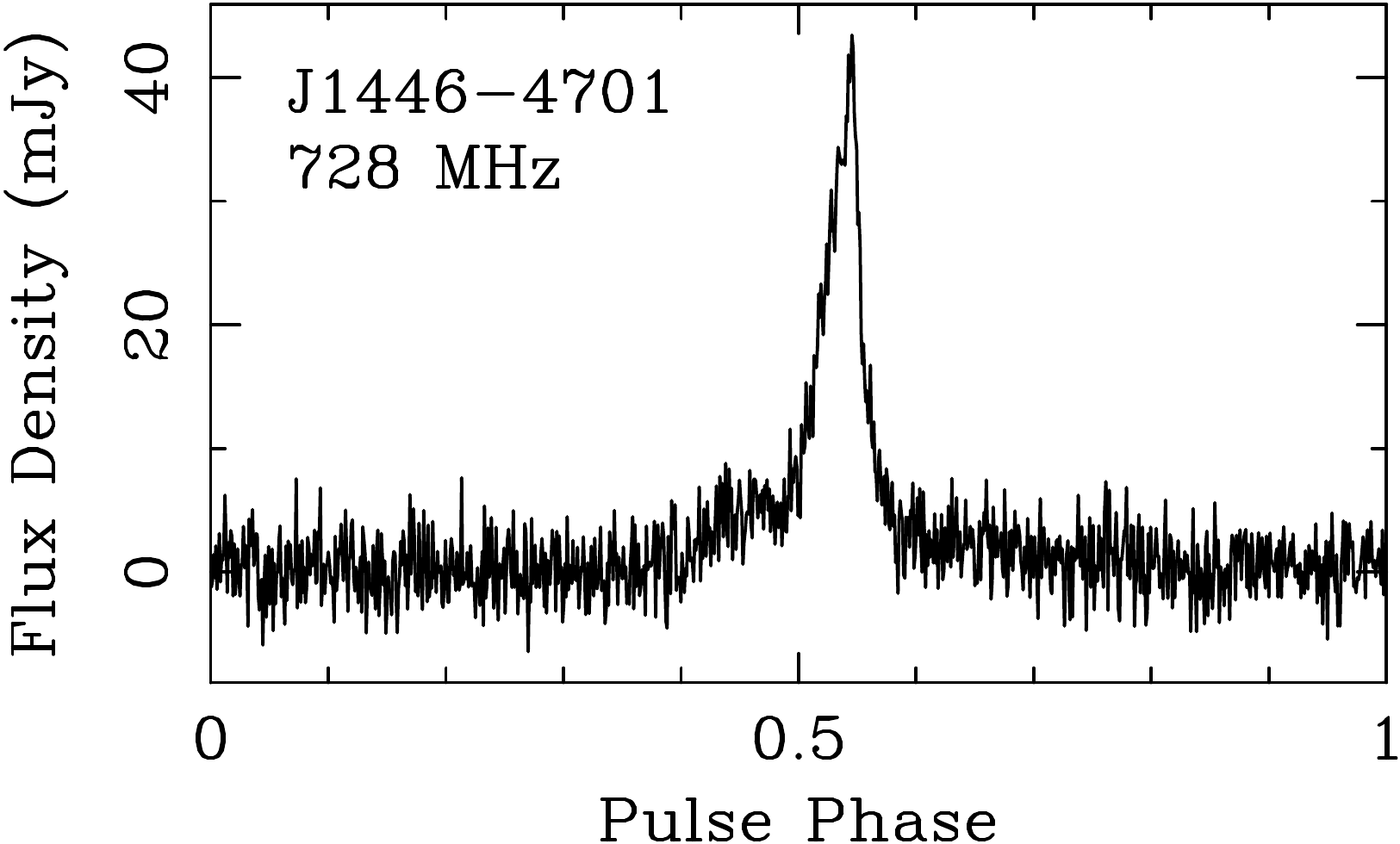}
                \caption{J1446-4701. $D=1.37 \pm 0.08$, Number of observation epoches: 19.}
                \label{fig:3}
        \end{subfigure}

          \end{tabular}
        \caption{Stokes parameter I of three MSPs plotted by PSRCHIVE software. Although J1446-4701 has less number of peaks but due to less $snr$, has higher fractal dimension. Fractality is more sensitive to $snr$ in comparison with number of peaks. Fig. \ref{fig:1} and Fig. \ref{fig:2} have the same observation conditions approximately, so the fractal dimension is demonstrating more complexity for J0711-6830. }
        \label{figure:PP}
\end{figure}

\begin{figure}
\centering
\includegraphics[scale=0.4]{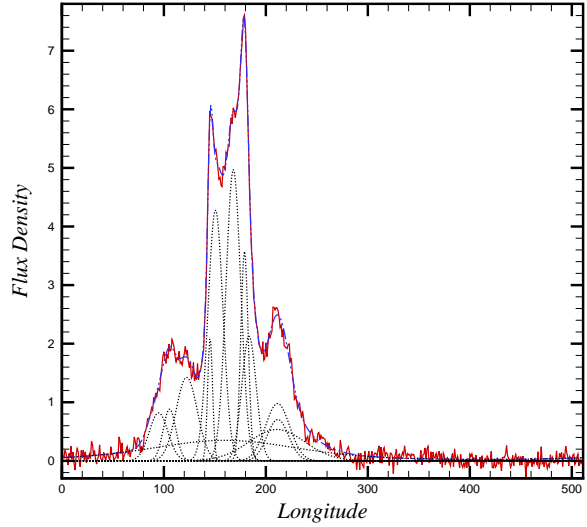} \\
\caption {Fitting Von Mises functions to observed Stokes parameter I of J0613-0200 at 3100 MHz. The Von Mises component are drawn in dotted curves, the sum of Von Mises components is dash-dotted curve which is a good representation of observed PPP (solid line).}
 \label{fig:paas}
\end{figure}

To do a fair comparison between parameters D or $\alpha$ of PPPs we need equivalent observational conditions, as much as possible. Despite fractal dimension of PPPs depends on pulse shape complexity, it also highly depends on quality of observation such as signal to noise ratio $(snr)$, number of observation epochs, PPPs resolution, etc (see Fig. \ref{figure:PP}). It is very hard to find PPPs with these equal conditions especially since we are working on two different samples namely normal and MSPs. Pulse profiles can be modeled by multiple Gaussian functions but recently most of authors uses Von Mises functions since they can fit the edges of components slightly better (\cite{von mise}). The Von Mises function for the angle $\theta$ is given by:
\begin{equation}
f(\phi)=I_0e^{\kappa\cos(\phi-\mu)-\kappa}
\label{VON}
\end{equation}
where $\kappa$ is the concentration, $I_0$ is the peak height and $\mu$ is the peak position. A part of PSRCHIVE software namely "paas" produces analytical pulse profiles by fitting Von Mises function to observed profiles (see Fig. \ref{fig:paas}) and we use it to produce denoised Stokes parameter I of MSPs and normal pulsars. Generally the number of Von Mises functions which are  needed for fitting is considered as the complexity of profile. However we believe fractal dimension can be a better measurement of complexity instead of number of fitted components. As a result, fitted pulse profiles enable us to do a fair comparison by removing noise from observed PPPs. The results of applying box counting and DFA methods on fitted Stokes parameter I are given in Tables \ref{tab:paas_MSPs} and \ref{tab:paas_reg}.

\section{Results and discussion}\label{res}
In this section we are intended to explain the results of fractal and Hurst exponent analyses on MSPs sample and seek for existence of any relationship between statistical parameters ($\alpha$ or D) of Stokes parameter I with pulsars' physical characteristics. Noticeably, we have used ATNF database for pulsars characteristics parameters (\cite{ATNF}). In addition we have chosen 24 normal pulsars randomly plus PPTA MSPs sample\footnote{We removed J1824-2452 from PPTA sample whose rotational period derivative is unknown up to now (\cite{Nature})} to make a comparison between MSPs and normal pulsars statistical parameters. The normal sample has been chosen from total 300 pulsars which their PPPs distributed by (\cite{Gould}) in European Pulsar Network (EPN) data archive. The normal and MSPs samples presented in Tables \ref{tab:paas_MSPs} and \ref{tab:paas_reg} have been observed at a common frequency of 1400\,MHz, then only the comparison of these samples in this bandwidth can be reliable.

\subsection{Complexity of Stokes parameter I profiles}
\begin{figure}
\centering
\includegraphics[scale=0.3]{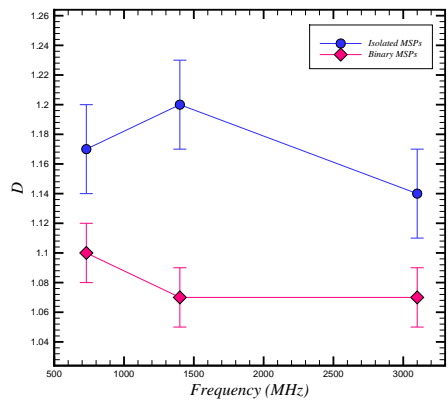} \\
\caption {Mean fractal dimension of binary and isolated MSPs Stokes parameter I. The KS-test result confirms that the difference of D between two sample is statistically meaningful.}
 \label{fig:D_Binary}
\end{figure}

\begin{figure}
\centering
\includegraphics[scale=0.3]{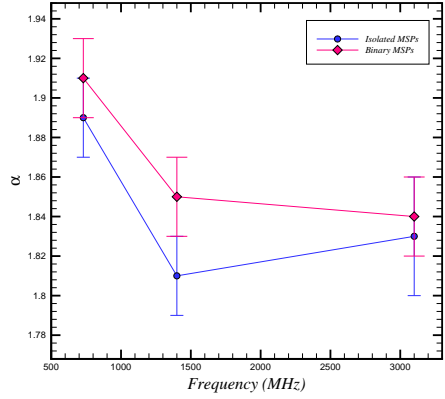} \\
\caption {Mean $\alpha$ exponent of binary and isolated MSPs Stokes parameter I. The KS-test result denied the existence of a meaningful difference for $\alpha$ exponent between two samples.}
 \label{fig:H_Binary}
\end{figure}

Comparing the mean values of fractal dimensions given in Tables \ref{tab:paas_MSPs} and \ref{tab:paas_reg} we conclude that MSPs PPPs exhibit more complexity in all frequencies including 1400\,MHz. Beside, normal pulsars PPPs fractal dimensions are not exceeding 1 and have not fractal properties. Furthermore, there are 7 isolated MSPs in our sample (J0711-6830, J1024-0719, J1730-2304, J1744-1134, J1832-0836, J1939+2134, J2124-3358). According to Table \ref{tab:binary}, Fig. \ref{fig:D_Binary}, and Fig. \ref{fig:H_Binary}, making a comparison between isolated and binary MSPs indicates that the isolated ones have more complex and less persistent PPPs in all three bands. In fact expect for PSR\,J1744-1134, fractal dimension of all isolated MSPs is higher than mean fractal dimension of binary MSPs in each band. To infer about any meaningful difference between normal and millisecond pulsars regarding to their fractal dimension and $\alpha$ exponent we use a non-parametric statistical test, namely two-sample Kolmogorov-Smirnov test (KS-test). In this test, null hypothesis is: the tested samples come from the same statistical distribution(\cite{Ks,newKS}). If a calculated P-value exceeds the critical value according to degrees of freedom we can infer that null hypothesis is rejected. The results of KS-test are given in Table. \ref{tab:KS} and shows the difference in fractal dimension is meaningful especially at 1408 MHz. However this test shows the difference in $\alpha$ exponent is not meaningful and two samples are from same distribution.
We believe that more complexity of isolated MSPs is more related to higher number of peaks and therefore more emission regions contribution in pulse profile.
\subsection{Emission regions}
Spatial interpretation of Hurst exponent states that for persistent PPPs if there is a rise (fall) in flux density in a particular phase point, it should be a rise (fall) in the neighborhood phase point statistically. In anti-persistent cases, a rise should be followed up by a fall and vice versa. According to Tables \ref{tab:paas_MSPs} and \ref{tab:paas_reg} all PPPs are non-stationary and Hurst exponent analysis indicates that the MSPs' PPPs are highly persistent however in normal pulsars there are persistent, random and anti-persistent cases. The very high values of $\alpha$ exponents in Table \ref{tab:paas_MSPs} means MSPs have very persistent pulse profiles. The very small and compact emission regions $(\rho)$ of MSPs can cause this high persistent property. In addition the density distribution of plasma in emission regions should be  persistent. It means in plasma of emission regions, in the vicinity of each maximum it should be maximum areas.
\subsection{Relationship with characteristic parameters}
Here we seek for any relationship between statistical properties of Stokes parameter I profiles (D and $\alpha$ exponent) with MSPs some characteristic parameters. To reach this goal, firstly we sort D or $\alpha$ exponent of MSPs sample according to a characteristic parameter. Then we split the sorted sample into two subsamples. One of them has low and the other one has high values of that characteristic parameter. Now we can apply KS-test to infer about any meaningful statistical difference between two subsamples. If there is any meaningful difference, we can say that characteristic parameter has a relationship with the statistical parameter (D or $\alpha$ exponent). The P-value of KS-test between D or $\alpha$ exponent of these two subsamples can reveal the existence of dependence between D or $\alpha$ exponent and that characteristic parameter. Table \ref{tab:MSP_D_KS} shows at least in one frequency, D  is dependent to: surface magnetic flux density $B_{surf}$, pulsar period $P$, time derivative of pulsar period $\dot{P}$ and the width of pulse at 50\% of peak. In addition according to Table \ref{tab:MSP_H_KS}, $\alpha$ exponent at least in one band is dependent to: pulsar period $P$ and the width of pulse at 50\% of peak as well, plus mean flux density at 1400 MHz.
Although isolated MSPs have lower luminosity than binary ones (\cite{Bailes,KramerI}), KS-test result exhibits there is no significant relationship between luminosity and complexity of PPPs.

\begin{acknowledgements}
The authors wish to thank Lawrence Toomey for his fantastic tutorials about installing TEMPO2 and PSRCHIVE pulsar timing software packages. Thanks to Professor Matthew Bailes for his useful recommendations about the complexity of pulse profiles. Thanks to Professor Andrew Lyne for sharing valuable information about the observation details of normal pulsars PPPs used in this paper.
\end{acknowledgements}

\newpage
\begin{table}
\caption{Fractal dimensions D and $\alpha$ exponents of fitted MSPs pulse profile of Stokes parameter I.}
\centering
\begin{minipage}{140mm}
\begin{tabular}{l*{6}{c}r}
\hline
PSR               &  & D &   &   & $\alpha$  &   \\

                  & 3100\,MHz & 1400\,MHz & 730\,MHz & 3100\,MHz  & 1400\,MHz & 730\,MHz \\
\hline
    J0437-4715 & 1.01  $\pm$ 0.07  & 1.04  $\pm$ 0.07  & 1.10  $\pm$ 0.09  & 1.89  $\pm$ 0.07  & 1.92  $\pm$ 0.06  & 1.88  $\pm$ 0.06 \\
    J0613-0200 & 1.15  $\pm$ 0.08  & 1.18  $\pm$ 0.08  & 1.12  $\pm$ 0.08  & 1.80  $\pm$ 0.08  & 1.75  $\pm$ 0.07  & 1.88  $\pm$ 0.07 \\
    J0711-6830 & 1.24  $\pm$ 0.07  & 1.38  $\pm$ 0.08  & 1.22  $\pm$ 0.08  & 1.94  $\pm$ 0.03  & 1.93  $\pm$ 0.03  & 1.93  $\pm$ 0.03 \\
    J1017-7156 & 1.03  $\pm$ 0.08  & 1.01  $\pm$ 0.07  & 1.03  $\pm$ 0.09  & 1.78  $\pm$ 0.08  & 1.68  $\pm$ 0.12  & 1.94  $\pm$ 0.12 \\
    J1022+1001 & 1.03  $\pm$ 0.08  & 1.07  $\pm$ 0.08  & 1.04  $\pm$ 0.08  & 1.87  $\pm$ 0.06  & 1.88  $\pm$ 0.07  & 1.86  $\pm$ 0.06 \\
    \\
    J1024-0719 & 1.12  $\pm$ 0.07  & 1.11  $\pm$ 0.07  & 1.15  $\pm$ 0.08  & 1.79  $\pm$ 0.06  & 1.79  $\pm$ 0.05  & 1.75  $\pm$ 0.06 \\
    J1045-4509 & 1.10  $\pm$ 0.08  & 1.11  $\pm$ 0.09  & 1.12  $\pm$ 0.08  & 1.95  $\pm$ 0.04  & 1.89  $\pm$ 0.09  & 1.97  $\pm$ 0.04 \\
    J1446-4701 & 1.03  $\pm$ 0.08  & 1.04  $\pm$ 0.08  & 1.02  $\pm$ 0.07  & 1.87  $\pm$ 0.10  & 1.83  $\pm$ 0.07  & 1.79  $\pm$ 0.09 \\
    J1545-4550 & 1.03  $\pm$ 0.08  & 1.06  $\pm$ 0.08  & 1.19  $\pm$ 0.08  & 1.82  $\pm$ 0.10  & 1.80  $\pm$ 0.10  & 1.98  $\pm$ 0.03 \\
    J1600-3053 & 1.07  $\pm$ 0.08  & 1.06  $\pm$ 0.08  & 1.07  $\pm$ 0.08  & 1.83  $\pm$ 0.09  & 1.83  $\pm$ 0.09  & 1.89  $\pm$ 0.05 \\
    \\
    J1603-7202 & 1.05  $\pm$ 0.07  & 1.09  $\pm$ 0.08  & 1.12  $\pm$ 0.09  & 1.77  $\pm$ 0.09  & 1.90  $\pm$ 0.05  & 1.94  $\pm$ 0.05 \\
    J1643-1224 & 1.09  $\pm$ 0.08  & 1.10  $\pm$ 0.08  & 1.13  $\pm$ 0.08  & 1.94  $\pm$ 0.08  & 1.97  $\pm$ 0.08  & 2.01  $\pm$ 0.04 \\
    J1713+0747 & 1.03  $\pm$ 0.08  & 1.03  $\pm$ 0.08  & 1.06  $\pm$ 0.08  & 1.79  $\pm$ 0.07  & 1.90  $\pm$ 0.08  & 1.91  $\pm$ 0.07 \\
    J1730-2304 & 1.14  $\pm$ 0.08  & 1.13  $\pm$ 0.09  & 1.10  $\pm$ 0.08  & 1.93  $\pm$ 0.04  & 1.73  $\pm$ 0.06  & 1.91  $\pm$ 0.04 \\
    J1744-1134 & 1.03  $\pm$ 0.08  & 1.03  $\pm$ 0.08  & 1.04  $\pm$ 0.09  & 1.90  $\pm$ 0.08  & 1.94  $\pm$ 0.07  & 1.93  $\pm$ 0.09 \\
    \\
    J1824-2452 & 1.05  $\pm$ 0.08  & 1.16  $\pm$ 0.08  & 1.21  $\pm$ 0.09  & 1.74  $\pm$ 0.08  & 1.64  $\pm$ 0.09  & 1.95  $\pm$ 0.07 \\
    J1832-0836 & 1.11  $\pm$ 0.08  & 1.26  $\pm$ 0.08  & 1.23  $\pm$ 0.07  & 1.82  $\pm$ 0.06  & 1.60  $\pm$ 0.07  & 1.92  $\pm$ 0.06 \\
    J1857+0943 & 1.23  $\pm$ 0.08  & 1.18  $\pm$ 0.08  & 1.26  $\pm$ 0.09  & 1.84  $\pm$ 0.07  & 1.97  $\pm$ 0.04  & 1.92  $\pm$ 0.06 \\
    J1909-3744 & 1.02  $\pm$ 0.08  & 1.01  $\pm$ 0.07  & 1.02  $\pm$ 0.08  & 1.69  $\pm$ 0.09  & 1.66  $\pm$ 0.09  & 1.85  $\pm$ 0.09 \\
    J1939+2134 & 1.08  $\pm$ 0.08  & 1.18  $\pm$ 0.08  & 1.14  $\pm$ 0.08  & 1.43  $\pm$ 0.17  & 1.72  $\pm$ 0.09  & 1.80  $\pm$ 0.07 \\
    \\
    J2124-3358 & 1.27  $\pm$ 0.08  & 1.29  $\pm$ 0.07  & 1.33  $\pm$ 0.09  & 1.97  $\pm$ 0.03  & 1.96  $\pm$ 0.02  & 1.99  $\pm$ 0.01 \\
    J2129-5721 & 1.13  $\pm$ 0.08  & 1.08  $\pm$ 0.09  & 1.07  $\pm$ 0.09  & 1.76  $\pm$ 0.07  & 1.89  $\pm$ 0.10  & 1.85  $\pm$ 0.09 \\
    J2145-0750 & 1.06  $\pm$ 0.07  & 1.10  $\pm$ 0.07  & 1.16  $\pm$ 0.08  & 1.93  $\pm$ 0.05  & 1.97  $\pm$ 0.04  & 1.88  $\pm$ 0.07 \\
    J2241-5236 & 1.02  $\pm$ 0.08  & 1.02  $\pm$ 0.08  & 1.02  $\pm$ 0.09  & 1.92  $\pm$ 0.08  & 1.81  $\pm$ 0.08  & 1.92  $\pm$ 0.07 \\
\hline
Mean & 1.10	$\pm$ 0.02 &	1.11 $\pm$	0.02 & 	1.12 $\pm$	0.02 & 	1.83 $\pm$	0.02 &	1.83 $\pm$	0.01 & 	1.90	$\pm$ 0.01 \\
\hline
\end{tabular}
\end{minipage}
\label{tab:paas_MSPs}
\end{table}

\newpage
\begin{table}
\caption{Fractal dimensions D and $\alpha$ exponents of fitted normal pulsars Stokes parameter I.}
\centering
\begin{minipage}{140mm}
\begin{tabular}{l*{6}{c}r}
\hline
PSR               &  & D &   &   & $\alpha$  &   \\

                  & 1642\,MHz & 1408\,MHz & 925\,MHz & 1642\,MHz & 1408\,MHz & 925\,MHz \\
\hline
    B0136+57 & 0.99  $\pm$0.07  & 1.00  $\pm$0.07  & 0.99  $\pm$0.07  & 1.53  $\pm$0.09  & 1.65  $\pm$0.09  & 1.56  $\pm$0.11 \\
    B0144+59 & 0.86  $\pm$0.07  & 0.97  $\pm$0.07  & 0.97  $\pm$0.07  & 1.28  $\pm$0.08  & 1.41  $\pm$0.08  & 1.56  $\pm$0.05 \\
    B0458+46 & 1.00  $\pm$0.08  & 0.99  $\pm$0.07  & 0.98  $\pm$0.07  & 1.67  $\pm$0.10  & 1.67  $\pm$0.14  & 1.66  $\pm$0.10 \\
    B0621-04 & 1.15  $\pm$0.07  & 0.94  $\pm$0.07  & 1.01  $\pm$0.07  & 1.98  $\pm$0.05  & 1.52  $\pm$0.12  & 1.45  $\pm$0.11 \\
    B0756-15 & 1.00  $\pm$0.07  & 0.94  $\pm$0.07  & 0.91  $\pm$0.07  & 1.47  $\pm$0.12  & 1.37  $\pm$0.10  & 1.51  $\pm$0.07 \\
    \\
    B0834+06 & 1.02  $\pm$0.07  & 0.96  $\pm$0.07  & 0.99  $\pm$0.07  & 1.74  $\pm$0.09  & 1.63  $\pm$0.12  & 1.55  $\pm$0.13 \\
    B0950+08 & 1.05  $\pm$0.08  & 1.04  $\pm$0.08  & 1.04  $\pm$0.08  & 1.82  $\pm$0.08  & 1.89  $\pm$0.06  & 1.86  $\pm$0.06 \\
    B1620-09 & 0.93  $\pm$0.07  & 0.95  $\pm$0.07  & 0.99  $\pm$0.07  & 1.45  $\pm$0.06  & 1.48  $\pm$0.05  & 1.51  $\pm$0.11 \\
    B1758-03 & 1.02  $\pm$0.07  & 1.00  $\pm$0.07  & 1.19  $\pm$0.08  & 1.54  $\pm$0.06  & 1.50  $\pm$0.09  & 1.44  $\pm$0.08 \\
    B1809-173 & 0.98  $\pm$0.07  & 0.96  $\pm$0.07  & 1.17  $\pm$0.08  & 1.65  $\pm$0.08  & 1.61  $\pm$0.09  & 1.76  $\pm$0.09 \\
    \\
    B1823-11 & 1.02  $\pm$0.07  & 1.00  $\pm$0.07  & 1.05  $\pm$0.07  & 1.50  $\pm$0.08  & 1.66  $\pm$0.07  & 1.68  $\pm$0.07 \\
    B1846-06 & 0.97  $\pm$0.07  & 0.93  $\pm$0.07  & 0.88  $\pm$0.07  & 1.50  $\pm$0.06  & 1.42  $\pm$0.05  & 1.44  $\pm$0.06 \\
    B1851-14 & 1.00  $\pm$0.07  & 1.01  $\pm$0.08  & 1.01  $\pm$0.07  & 1.62  $\pm$0.10  & 1.75  $\pm$0.08  & 1.65  $\pm$0.08 \\
    B1900+01 & 0.96  $\pm$0.07  & 0.99  $\pm$0.07  & 0.92  $\pm$0.07  & 1.45  $\pm$0.13  & 1.58  $\pm$0.08  & 1.40  $\pm$0.11 \\
    B1913+167 & 1.11  $\pm$0.07  & 1.02  $\pm$0.07  & 1.28  $\pm$0.07  & 1.56  $\pm$0.10  & 1.62  $\pm$0.09  & 1.99  $\pm$0.02 \\
    \\
    B1924+16 & 0.97  $\pm$0.07  & 0.99  $\pm$0.07  & 0.98  $\pm$0.07  & 1.59  $\pm$0.08  & 1.56  $\pm$0.10  & 1.51  $\pm$0.09 \\
    B1929+10 & 1.02  $\pm$0.08  & 1.03  $\pm$0.08  & 1.03  $\pm$0.08  & 1.68  $\pm$0.10  & 1.75  $\pm$0.08  & 1.74  $\pm$0.09 \\
    B1935+25 & 1.14  $\pm$0.08  & 1.06  $\pm$0.08  & 1.20  $\pm$0.09  & 1.32  $\pm$0.11  & 1.42  $\pm$0.09  & 1.43  $\pm$0.10 \\
    B1953+50 & 0.98  $\pm$0.07  & 0.98  $\pm$0.07  & 0.95  $\pm$0.07  & 1.42  $\pm$0.12  & 1.52  $\pm$0.07  & 1.42  $\pm$0.06 \\
    B2021+51 & 1.01  $\pm$0.07  & 1.01  $\pm$0.07  & 1.01  $\pm$0.07  & 1.71  $\pm$0.10  & 1.73  $\pm$0.10  & 1.68  $\pm$0.11 \\
    \\
    B2053+36 & 0.93  $\pm$0.07  & 1.00  $\pm$0.07  & 0.95  $\pm$0.08  & 1.37  $\pm$0.04  & 1.63  $\pm$0.10  & 1.43  $\pm$0.09 \\
    B2148+52 & 0.95  $\pm$0.07  & 1.01  $\pm$0.08  & 1.03  $\pm$0.08  & 1.59  $\pm$0.07  & 1.65  $\pm$0.08  & 1.62  $\pm$0.10 \\
    B2227+61 & 1.01  $\pm$0.08  & 1.01  $\pm$0.08  & 1.09  $\pm$0.08  & 1.37  $\pm$0.09  & 1.66  $\pm$0.08  & 1.61  $\pm$0.07 \\
    B2255+58 & 0.99  $\pm$0.07  & 1.02  $\pm$0.07  & 1.01  $\pm$0.07  & 1.53  $\pm$0.09  & 1.77  $\pm$0.09  & 1.69  $\pm$0.08 \\
\hline
Mean & 1.00	$\pm$ 0.01 &	0.99 $\pm$	0.01 & 1.03	$\pm$ 0.02	& 1.55	$\pm$ 0.02 & 1.60	$\pm$	0.02 &	1.59	$\pm$	0.02 \\

\hline
\end{tabular}
\end{minipage}
\label{tab:paas_reg}
\end{table}

\newpage
\begin{table}
\caption{Mean values of D and $\alpha$ exponent for Stokes parameter I of isolated and binary MSPs.}
\centering
\begin{minipage}{140mm}
\begin{tabular}{l*{6}{c}r}
\hline
            &  & D &   &   & $\alpha$  &   \\
                  & 3100\,MHz & 1400\,MHz & 730\,MHz & 3100\,MHz  & 1400\,MHz & 730\,MHz \\
\hline
Binary  &  1.07  $\pm$ 0.02  & 1.07  $\pm$ 0.02  & 1.10  $\pm$ 0.02  & 1.84  $\pm$ 0.02  & 1.85  $\pm$ 0.02  & 1.91  $\pm$ 0.02 \\
Isolated   &  1.14  $\pm$ 0.03  & 1.20  $\pm$ 0.03  & 1.17  $\pm$ 0.03  & 1.83  $\pm$ 0.03  & 1.81  $\pm$ 0.02  & 1.89  $\pm$ 0.02 \\
\hline
\end{tabular}
\end{minipage}
\label{tab:binary}
\end{table}

\newpage
\begin{table}
\caption{P-values of KS-test after splitting MSPs sample into isolated and binary pulsars.}
\centering
\begin{minipage}{140mm}
\begin{tabular}{l*{6}{c}r}
\hline
Samples           &  & D &   &   & $\alpha$  &   \\
                  & 3100\,MHz & 1400\,MHz & 730\,MHz & 3100\,MHz  & 1400\,MHz & 730\,MHz \\
\hline
Isolated-Binary   &  0.07011	&0.00469	&0.08702	&0.60384	&0.37580	&0.77984 \\
\hline
\end{tabular}
\end{minipage}
\label{tab:KS}
\end{table}

\begin{table}
\caption{P-values of KS-test between D with characteristic parameters in different bands.}
\centering
\begin{minipage}{140mm}
\begin{tabular}{c*{6}{c}r}
\hline
 Band(MHz)  &  $B_{surf}$  &    P  & $\dot{P}$  &  W50   \\
\hline
3100 & 0.862        & 0.241 & 0.551      &  0.026 \\
1400 & 0.083        & 0.026 & 0.083      &  0.001 \\
730  & 0.026        & 0.026 & 0.083      &  0.026 \\
\hline
\end{tabular}
\end{minipage}
\label{tab:MSP_D_KS}
\end{table}
\begin{table}
\caption{P-values of KS-test between $\alpha$ exponent with characteristic parameters in different bands.}
\centering
\begin{minipage}{140mm}
\begin{tabular}{c*{6}{c}r}
\hline
 Band(MHz)  &   P    & S1400 &   W50   \\
\hline
3100 & 0.029  & -- &  0.040 \\
1400 & 0.029  & 0.111 &  0.260 \\
730  & 0.616  & -- &  0.616 \\
\hline
\end{tabular}
\end{minipage}
\label{tab:MSP_H_KS}
\end{table}


\clearpage
\begin{thebibliography}{}


\bibitem[\protect\citeauthoryear{Bailes et al}{1997}]{Bailes} Bailes M., Johnston S., Bell J. F., Lorimer D. R., Stappers B. W., Manchester R. N., ... \& Gaensler B. M., 1997, Springer Berlin Heidelberg.



\bibitem[\protect\citeauthoryear{Chandrasekhar}{1960}]{Chandrasekhar} Chandrasekhar S., 1960, Dover Publications, New York, 1960, ISBN 0-486-60590-6.


\bibitem[\protect\citeauthoryear{Chen et al.}{2002}]{Chen} Chen Z., Ivanov P. C., Hu K., $\&$ Stanley H. E., 2002, physical Review E, 65(4), 041107.


\bibitem[\protect\citeauthoryear{Dai et al.}{2015}]{PPTA} Dai S., Hobbs G., Manchester R. N., Kerr M., Shannon R. M., van Straten W., ... $\&$ Zhu, X. J., 2015, Monthly Notices of the Royal Astronomical Society, 449(3), 3223-3262.




\bibitem[\protect\citeauthoryear{Feder}{1988}]{Feder} Feder J., 1988, Plenum.


\bibitem[\protect\citeauthoryear{Gil \& Krawczyk}{1997}]{Gil} Gil J., \& Krawczyk A., 1997, Monthly Notices of the Royal Astronomical Society, 285(3), 561-566.


\bibitem[\protect\citeauthoryear{Gneiting \& Schlather}{2004}]{Gneiting} Gneiting T., Schlather M, 2004, SIAM review, 46(2), 269-282.

\bibitem[\protect\citeauthoryear{Gould}{1998}]{Gould} Gould D. M., \& Lyne A. G. , 1998, Monthly Notices of the Royal Astronomical Society, 301(1), 235-260.

\bibitem[\protect\citeauthoryear{Hankins \& Eilek}{2007}]{hankin} Hankins T. H., \& Eilek J. A. , 2007,The Astrophysical Journal, 670(1), 693.

\bibitem[\protect\citeauthoryear{Hardstone et al.}{2012}]{Hardstone} Hardstone R., Poil S. S., Schiavone G., Jansen, R. Nikulin V. V., Mansvelder H. D., $\&$ Linkenkaer-Hansen K., 2012, Frontiers in physiology, 3.

\bibitem[\protect\citeauthoryear{Hu et al.}{2001}]{Hu} Hu K., Ivanov P. C., Chen Z., Carpena P., $\&$ Stanley H. E., 2001, Physical Review E, 64(1), 011114.


\bibitem[\protect\citeauthoryear{Kantelhardt et al.}{2002}]{mfDFA} Kantelhardt, J. W. Zschiegner, S. A. Koscielny-Bunde, E. Havlin, S. Bunde, A. \& Stanley H. E. (2002), Physica A: Statistical Mechanics and its Applications, 316(1), 87-114.

\bibitem[\protect\citeauthoryear{Karastergiou \& Johnsto}{2007}]{Karastergiou} Karastergiou A., \& Johnsto, S. (2007), Monthly Notices of the Royal Astronomical Society, 380(4), 1678-1684.


\bibitem[\protect\citeauthoryear{Kobayashi et al.}{2011}]{japoni} Kobayashi N., Yamazaki Y., Kuninaka H., Katori M., Matsushita M., Matsushita S., \& Chiang L. Y., 2011, Journal of the Physical Society of Japan, 80(7), 074003.


\bibitem[\protect\citeauthoryear{Kolmogorov}{1933}]{Ks} Kolmogorov A. N., 1933, Na.


\bibitem[\protect\citeauthoryear{Kramer et al.}{1998}]{KramerI} Kramer M., Xilouris K. M., Lorimer D. R., Doroshenko O., Jessner A., Wielebinski R., ... \& Camilo F., 1998, Spectra, pulse shapes, and the beaming fraction. The Astrophysical Journal, 501(1), 270.


\bibitem[\protect\citeauthoryear{Lyne \& Manchester}{1988}]{lyne} Lyne A. G., \& Manchester R. N., 1988, Monthly Notices of the Royal Astronomical Society, 234(3), 477-508.

\bibitem[\protect\citeauthoryear{Manchester}{1995}]{manchester2} Manchester R. N., 1995, Journal of Astrophysics and Astronomy, 16(2), 107-117.


\bibitem[\protect\citeauthoryear{Manchester et al.}{2005}]{ATNF}  Manchester R. N., Hobbs G. B., Teoh A. $\&$ Hobbs M., 2005, 129, The Astronomical Journal, 129(4), 1993.

\bibitem[\protect\citeauthoryear{Maragos \& Sun}{1993}]{fractal_signal} Maragos P., \& Sun F. K., 1993, IEEE Transactions on signal Processing, 41(1), 108-121.

\bibitem[\protect\citeauthoryear{Matos et al.}{2004}]{Matos} Matos J. M. O., de Moura E. P., Kruger S. E., Rebello J. M. A., 2004, Chaos, Solitons $\&$ Fractals, 19(1), 55-60.

\bibitem[\protect\citeauthoryear{Movahed et al.}{2006}]{movahed} Movahed M. S., Jafari G. R., Ghasemi F., Rahvar S., \& Tabar M. R. R., 2006, Journal of Statistical Mechanics: Theory and Experiment, 2006(02), P02003.


\bibitem[\protect\citeauthoryear{Papitto et al.}{2013}]{Nature} Papitto A., Ferrigno C., Bozzo E., Rea N., Pavan L., Burderi L., $... \&$ Wong, G. F., 2013, Nature, 501(7468), 517-520.

\bibitem[\protect\citeauthoryear{Peng et al.}{1994}]{Peng} Peng C. K., Buldyrev S. V., Havlin S., Simons M., Stanley H. E., $\&$ Goldberger A. L., 1994, Physical Review E, 49(2), 1685.

\bibitem[\protect\citeauthoryear{Racine}{2011}]{Racine} Racine R., 2011, Zurich: Mosaic Group.

\bibitem[\protect\citeauthoryear{Rankin}{1983}]{Rankin} Rankin J. M., 1983, The Astrophysical Journal, 274, 333-368.


\bibitem[\protect\citeauthoryear{Rypdal}{2012}]{Rypdal} Rypdal K., \& Rypdal M., 2012, In Multi-scale Dynamical Processes in Space and Astrophysical Plasmas (pp. 227-232). Springer Berlin Heidelberg.

\bibitem[\protect\citeauthoryear{Stephens}{1970}]{newKS} Stephens, M. A., 1970, Journal of the Royal Statistical Society. Series B (Methodological), 115-122.

\bibitem[\protect\citeauthoryear{Tauris \& Van Den Heuvel}{2006}]{evolution} Tauris T. M., \& Van Den Heuvel E. P. J., 2006, Compact stellar X-ray sources, 129,1, 623-665.

\bibitem[\protect\citeauthoryear{Van Straten et al.}{2010}]{stokes} Van Straten W., Manchester R. N., Johnston S., \& Reynolds J. E., 2010, ublications of the Astronomical Society of Australia, 27(1), 104-119.


\bibitem[\protect\citeauthoryear{Weron}{2002}]{Weron} Weron R., 2002, Physica A: Statistical Mechanics and its Applications, 312(1), 285-299.

\bibitem[\protect\citeauthoryear{Weltevrede \& Johnston}{2002}]{von mise} Weltevrede P., \& Johnston S., 2008, Monthly Notices of the Royal Astronomical Society, 391(3), 1210-1226.


\end{thebibliography}
\end{document}